\documentclass[a4paper,10pt,conference]{IEEEtran}
\usepackage{times} 
\usepackage[nocompress]{cite}
\usepackage{amsmath,amssymb,amsfonts}
\usepackage{algorithmic}
\usepackage{float}
\usepackage{graphicx}
\usepackage{textcomp}
\usepackage{url}
\usepackage{tabularx}
\usepackage{comment}
\usepackage{array} 
\usepackage{makecell}
\usepackage{multirow}
\usepackage{booktabs}
\usepackage[colorlinks,urlcolor=blue,linkcolor=blue,citecolor=blue]{hyperref}
\usepackage[dvipsnames,table]{xcolor}
\definecolor{LightCyan}{rgb}{0.88,1,1}
\usepackage{silence}
\usepackage[justification=centering]{caption}
\usepackage{balance}
\usepackage{graphicx} 
\usepackage{adjustbox} 

\begin{document}

\title{Navigating MLOps: Insights into Maturity, Lifecycle, Tools, and Careers}

\author{\IEEEauthorblockN{Jasper Stone\IEEEauthorrefmark{1},  Raj Patel\IEEEauthorrefmark{2},
Farbod Ghiasi\IEEEauthorrefmark{3},
Sudip Mittal\IEEEauthorrefmark{4},
Shahram Rahimi\IEEEauthorrefmark{5}
}
 
\IEEEauthorblockA{Dept. of Computer Science and Engineering,
Mississippi State University\\
\{jws819\IEEEauthorrefmark{1},
rkp88\IEEEauthorrefmark{2},
fg289\IEEEauthorrefmark{3}\}@msstate.edu,
mittal@cse.msstate.edu\IEEEauthorrefmark{4}
rahimi@cse.msstate.edu\IEEEauthorrefmark{5}}
}

\maketitle

\begin{abstract}
The adoption of Machine Learning Operations (MLOps) enables automation and reliable model deployments across industries. However, differing MLOps lifecycle frameworks and maturity models proposed by industry, academia, and organizations have led to confusion regarding standard adoption practices. This paper introduces a unified MLOps lifecycle framework, further incorporating Large Language Model Operations (LLMOps), to address this gap. Additionally, we outlines key roles, tools, and costs associated with MLOps adoption at various maturity levels. By providing a standardized framework, we aim to help organizations clearly define and allocate the resources needed to implement MLOps effectively.
\end{abstract}

\begin{IEEEkeywords}
Automated Machine Learning, Continuous Integration, DevOps, Large Language Models, Monitoring
\end{IEEEkeywords}

\section{Introduction}
Consumers increasingly expect AI-powered features in their daily interactions, ranging from weather forecasting to search and chat functionalities. Machine learning engineers and software developers must deploy these applications in scalable production-ready environments to ensure their responsiveness and availability. A study by Business Insights (2025) projects a 43\% growth in the MLOps market within five years\cite{fortunebusinessinsights_mlops_market}. However, the study by an algorithmia analysis conducted in 2020 found that 55\% of businesses using machine learning actively had not yet produced a model, and 18\% required more than 90 days or more\cite{algorithmia}. These prolonged deployment timelines illustrate the additional complexities associated with machine learning compared to traditional software and highlight the growing market demand for effective deployment solutions\cite{Lima2022}.

The continuous maintenance of machine learning models after deployment is necessary to address performance issues caused by data drift, unlike typical software\cite{John2021}. Reproducibility is likewise difficult since small deviations in data or algorithms can produce inconsistent results, complicating model validation. Debugging and updates are difficult without automated deployment due to the differences in configurations across environments\cite{Hewage2022}. Monitoring of model accuracy and latency along with periodic retraining with new data is crucial to staying current\cite{Shan2024}. These complexities affect manufacturing, business intelligence, aerospace and transportation among other industries.

Machine Learning Operations (MLOps) has emerged as a framework for deploying incremental machine learning improvements in a consistent and reliable manner. Despite the rise in MLOps adoption, there is no consensus among industries on what a comprehensive MLOps lifecycle framework entails \cite{Symeonidis2022} or on incorporating recent advancements in Large Language Models (LLMs) in the existing MLOps lifecycle. Given the recent rise in the popularity of LLMs, teams will be faced with the decision to train their own model or use an existing LLM. What criteria should be used to prefer one path over the other and how does this choice affect the MLOps lifecycle? Are there additional skills or technology needed if the team chooses to use a LLM?

As organizations embark on their MLOps journey, an evaluation of their MLOps maturity should be performed. MLOps maturity is defined as the qualitative assessment of people, processes, and technology involved in deploying and monitoring ML applications. At it's lowest level, maturity often involves manual creation and deployment, resulting in numerous challenges and limited observability. As maturity progresses, the deployment processes become more reliable, repeatable, and scalable. Highlighting the necessary roles and skill requirements allows teams to identify gaps and strategically plan for skill development and labor costs. Furthermore, identifying essential MLOps tools enables teams to align their skills and resource allocation.

This paper seeks to address questions about the lifecycle framework and maturity, providing insight into the personnel, tools, and planning required for successful implementation of MLOps. Therefore, our main contributions in this paper include:

\begin{itemize}
    \item \textbf{Consolidated MLOps Lifecycle framework:} We proprose a novel MLOps lifecycle framework combining best practices from academia and industry.
    \item \textbf{Inclusion of LLMOps:} Our proposed MLOps lifecycle framework includes LLMOps and the guidance for evaluating whether to either train your own LLM model or use an existing one.
    \item \textbf{Roles:} We provide a comprehensive list of roles involved throughout the MLOps lifecycle, from ideation through deployment and monitoring.
    \item \textbf{Tools:} An overview of tools supporting machine learning system development, including associated costs and alignment with MLOps lifecycle framework phases.
\end{itemize}

\section{MLOps Lifecycle and The Taxonomy} \label{sec:mlops_lifecycle}
Machine Learning Operations incorporates Software DevOps principles like Continuous Integration/Continuous Deployment (CI/CD) into machine learning workflows to improve collaboration, accelerate deployment, and ensure model reproducibility. Introduced in 2015 \cite{mlops-coin}, MLOps still lacks a standardized lifecycle, with various maturity models offering different stages and perspectives. Table \ref{tab:mlops_maturity_lvl} shows some maturity levels proposed by tech giants and researchers. These models generally begin with manual ML processes and evolve to full automation, covering data collection, model development, versioning, delivery, monitoring, and retraining, often with AutoML \cite{Symeonidis2022}.

\renewcommand{\arraystretch}{1.2}
\begin{table}[h!]
    \centering
    \begin{tabular}{|p{2.6cm}|p{5.2cm}|}
        \hline
        \rowcolor{LightCyan} {\bf Company/Paper} &  {\bf Maturity Level Proposed(s)}\\
        \hline
        \multirow{3}{*}{Google \cite{google-mlops}}
        & Level 0: Manual Process \\
        & Level 1: ML Pipeline Automation \\
        & Level 2: Automated Training \\
        \hline
        \multirow{5}{*}{Microsoft \cite{google-mlops}}
        & Level 0: No MLOps \\
        & Level 1: DevOps but No MLOps \\ 
        & Level 2: Automated Training \\
        & Level 3: Automated Model Deployment \\ 
        & Level 4: Full MLOps Automated Operations \\
        \hline
        \multirow{4}{*}{Amazon \cite{Amazon-mlops}}
        & Level 1: Initial Phase  \\ 
        & Level 2: Repeatable Phase \\ 
        & Level 3: Reliable Phase \\ 
        & Level 4: Scalable Phase \\
        \hline
        \multirow{4}{*}{IBM \cite{IBM-mlops}} 
        & Level 0: No MLOps  \\ 
        & Level 1: ML Pipeline Automation \\ 
        & Level 2: CI/CD Pipeline Integration \\ 
        & Level 3: Advanced MLOps \\
        \hline
        \multirow{4}{*}{Meenu Mary et al. \cite{John2021}} 
        & Level A: Automated Data Collection \\
        & Level B: Automated Model Deployment \\ 
        & Level C: Semi-automated Model Monitoring \\ 
        & Level D: Fully-automated Model Monitoring \\
        \hline
        \end{tabular}
    \caption{Recommended MLOps maturity levels in industry and research. Higher levels imply a higher level of maturity.}
    \label{tab:mlops_maturity_lvl}
\end{table}

\begin{figure*}[!htb]
        \centering
        \includegraphics[scale=0.328]{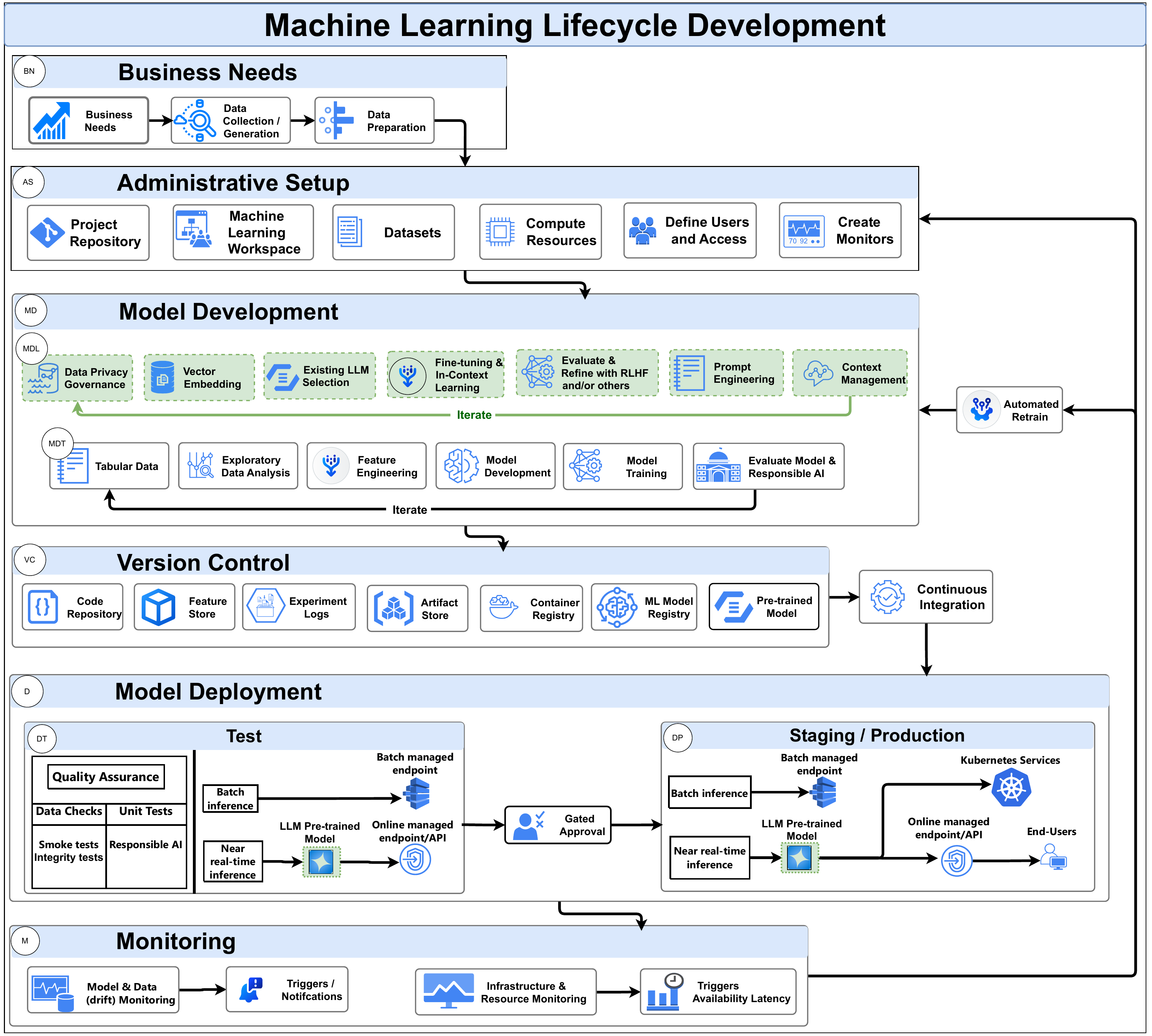}
        \caption{The consolidated MLOps lifecycle begins at the top-left and progresses left to right within each phase. The forest-green highlighted pipeline represents the LLMOps development cycle and is not complete as it needs other phases as well. Each completed phase leads downward to monitoring, whose feedback loops back into administrative setup and model development, enabling iterative improvement.}
        \vspace {-0.3 cm}
        \label{fig:MLOPS}
    \end{figure*}

As MLOps process matures, aligning them with business needs becomes paramount to ensure that machine learning initiatives deliver tangible value. Figure \ref{fig:MLOPS} illustrates a MLOps maturity, where the highest level of automation and monitoring is implemented. The following description provides a detailed overview of our MLOps lifecycle framework, with terms in italics corresponding to those shown in Figure \ref{fig:MLOPS}.

        \textit{Business Needs (BN)} take first priority, as MLOps is driven by the need to align machine learning initiatives with business objectives to address strategic challenges \cite{tolpegin2020data}. Data scientists set clear Objectives and Key Results (OKRs) focused on decision-making processes and critical operations \cite{testi2022mlops}. Robust ML models are crucial for positive business outcomes \cite{pitropakis2019taxonomy} \cite{yerlikaya2022data}. As organizations scale from proof-of-concept to operational machine learning, MLOps pipelines enable continuous retraining, deployment, and scalability \cite{makinen2021needs}. Once business needs are identified, the next essential step is to collect and/or generate the data required to train models that support these objectives. Following business needs, \textit{Data Collection/Generation} is central to machine learning. Data is sourced from proprietary databases, online repositories, sensors, or user-generated content \cite{yerlikaya2022data}, using methods such as web scraping, APIs, or crowdsourcing platforms such as Amazon Mechanical Turk \cite{tolpegin2020data} \cite{makinen2021needs}. When data is limited, augmentation or transfer learning from pre-trained models may be employed \cite{kreuzberger2023machine, zhang2022conceptualizing}. Data management systems oversee versioning, lineage, governance, and quality control, with collaboration tools such as JIRA and Slack that facilitate communication \cite{makinen2021needs}. After gathering the necessary data, proper preparation techniques are employed to ensure the quality and consistency of inputs used for model development.
        
        In \textit{Data Preparation}, data is manipulated to improve model performance. This involves cleaning to handle missing values, errors, or duplicates  \cite{kumara2022requirements}, and normalizing input values to ensure consistency for algorithm efficiency \cite{yerlikaya2022data, kumara2022requirements}. Data augmentation techniques, such as rotation and scaling, are applied to artificially expand datasets \cite{kreuzberger2023machine}. Integration of multiple data sources and careful versioning ensure reproducibility and consistency throughout the model lifecycle \cite{makinen2021needs}. Once the business needs have been identified and data collected, we move into administrative setup to create the environments for the team to collaborate.
        
        Establishing a robust administrative framework in the \textit{Administrative \& Setup (AS)} phase is crucial for managing and streamlining MLOps workflows. This involves setting up project and versioning repositories, machine learning workspaces, defining user roles and access controls, and scaling hardware to accommodate varying demands. \textit{Project Repositories} are used to facilitate collaborative work on code, models, and data. This ensures that team members can collaborate, track changes, experiment and backup their work. Following this, machine learning workspaces are configured. \textit{Machine Learning Workspaces} are created for model development. These workspaces leverage platforms like MLflow\cite{mlflow}, Kubeflow\cite{kubeflow}, JupyterHub, and Databricks\cite{databricks} to centralize tools, datasets, and computational resources, streamlining workflows and boosting team productivity. After establishing workspaces, environments for data ingestion and transformation are set up using tools like Labelbox or Amazon SageMaker Ground Truth \cite{noauthor_undated-jr}. Accurate labeling is essential for effective model training. Following data labeling, datasets are stored and versioned. \textit{Datasets} are structured and managed to ensure data integrity, confidentiality, and compliance with data protection regulations. Proper dataset management supports the project lifecycle and ensures data quality for model training. 
        
        Provisioning \textit{Compute Resources} (CPU, GPU, TPU) is essential for model training and deployment. This includes both cloud-based and on-premises resources to meet computational demands. Adequate compute resources are crucial for efficient model experimentation and deployment. \textit{Defining Users and Access} involves setting permissions to ensure secure and efficient collaboration. Tools such as Identity and Access Management (IAM) are employed to manage access controls, ensuring security and data integrity \cite{noauthor_undated-fh}. Administrators \textit{Create Monitors} using tools like Prometheus and Elasticsearch, Logstash, Kibana (ELK) Stacks, are established to track system performance and detect anomalies. Effective monitoring is critical for maintaining system health and ensuring smooth operation throughout the MLOps lifecycle framework. Once the administrative setup is in place, it is crucial to implement robust version control mechanisms to track changes and ensure reproducibility throughout the pipeline.

        
        \begin{figure*}[!htb]
            \centering
            \includegraphics[width=\textwidth]{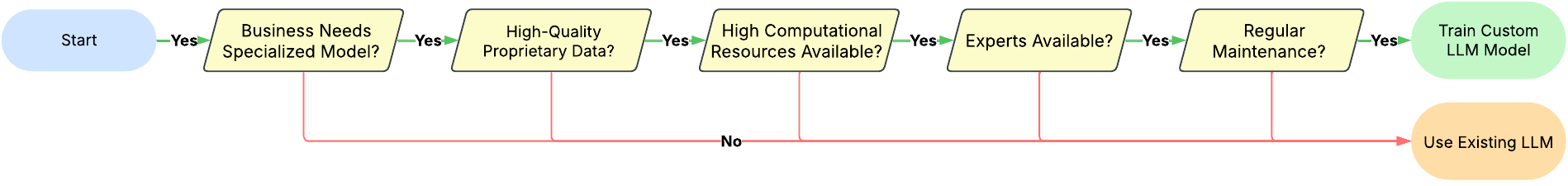}
            \caption{Recommended decision process for training your own custom LLM model vs using an existing LLM.}
            \label{fig:llm-decision}
            \vspace{-0.5 cm}
        \end{figure*}
        
        During \textit{Model Development (MD)}, a decision must be made to either leverage an existing large language model (LLM) or train a custom model. Unless your application needs a specialized model and have access to ample high-quality data, significant computational resources, specialized machine learning expertise, and can continuously maintain the model, using an existing LLM is generally preferable, as illustrated in Figure \ref{fig:llm-decision}. Utilizing a pre-existing model allows your team to build upon previous innovations, significantly accelerating deployment. The \textit{(MDL)} stage within model development, highlighted in forest green with dotted lines in Figure \ref{fig:MLOPS}, encompasses several key tasks: \textit{Data Privacy Governance} focuses on eliminating any data that could violate privacy or governance regulations, such as GDPR and CCPA \cite{lakeFS_llmops}. \textit{Vector Embedding} involves storing high-dimensional data efficiently as vectors, particularly beneficial for LLMOps tasks \cite{lakeFS_llmops}. \textit{Existing LLM Selection} process requires identifying suitable models such as LLaMA, GPT-J, Flan-T5 or Stable Diffusion \cite{lakeFS_llmops} to address specific business tasks. \textit{Fine Tuning and In-Context Learning} involves adapting existing LLMs via techniques such as zero-shot prompting, where only the prompt is given, or few-shot prompting, where examples are provided for the model's context. Fine-tuning adjusts weights in the model's final layers, resulting in a smaller, optimized language model ready for deployment. This approach enhances accuracy, reduces latency, and improves cost efficiency. One such example is to train a model to generate sarcastic responses \cite{Biswas2022}. In-context learning (ICL) involves inserting context data directly into the LLM prior to the user's prompt, enabling the model to reason effectively. Practical examples include document summarization or embedding data chunks from retrieval-augmented generation (RAG) systems.
        
        \textit{Evaluate \& Refine with RLHF and/or others} entails tuning parameters such as temperature, max-tokens, top-k, top-p, frequency penalty, and presence penalty, each affecting creativity, response length, and predictability. Model evaluation in LLMOps typically employs intrinsic metrics like ROUGE, BERT, and BLEU scores, assessing how closely outputs match reference answers. Human evaluation involves experts or crowdsourced evaluators examining LLM performance in specific scenarios. Task-specific benchmarks like GLUE or SuperGLUE offer standardized tasks and metrics for comprehensive assessment \cite{medium_llmops}. Additionally, unit and integration testing tools such as DeepEval \cite{DeepEval}, BLEU \cite{databricks_llmops}, and ROUGE \cite{databricks_llmops} verify model performance. Reassessing Fairness, Accountability, Transparency, and Ethics (FATE) \cite{Memarian2023} ensures that fine-tuning and prompt engineering have not introduced bias or other undesirable outcomes. Occasionally, models are further refined through reinforcement learning with human feedback (RLHF) \cite{databricks_llmops}. \textit{Prompt Engineering} is the strategic crafting of prompts to guide an LLM toward desired outputs. Techniques include zero-shot, few-shot, chain-of-thought, and prompt-chaining \cite{Shan2024} \cite{Chen2024}. \textit{Context Management} involves effectively managing relevant contextual information to ensure the accuracy and relevance of LLM responses prior to deployment.

        The \textit{(MDT)} section of model development, shown in solid lines, focuses on iteratively building and refining machine learning models. This includes working with different data types, algorithm selection, and feature engineering. The first step is to ensure the data is in a format that facilitates machine learning. \textit{Tabular Data} structures are well-suited for various machine learning algorithms, revealing hidden patterns through de-normalization \cite{makinen2021needs}. Despite potential data redundancy, this format enhances data retrieval and supports effective pre-processing, feature engineering, and visualization, crucial for improving model performance and facilitating \textit{Exploratory Data Analysis (EDA)} \cite{testi2022mlops}. EDA involves using visualization tools like histograms and scatter plots to explore correlations, patterns, and anomalies within the data \cite{John2021}. By summarizing key characteristics through statistical methods, EDA helps detect outliers and guide feature selection thus improving model accuracy \cite{Lima2022}. \textit{Feature Engineering} transforms raw data into meaningful inputs for models, employing techniques like  Principal Component Analysis (PCA) for dimensionality reduction and hyper-parameter tuning for model optimization \cite{makinen2021needs} \cite{kreuzberger2023machine}. Creating new features from existing data through techniques like polynomial features enhances model robustness and accuracy \cite{kreuzberger2023machine}.
        
        \textit{Model Development} involves selecting the appropriate algorithm, such as linear regression, decision trees, or neural networks, depending on the data and problem context \cite{testi2022mlops} \cite{kreuzberger2023machine}. Ensuring model interpretability and explainability is crucial for building stakeholder trust, while evaluation metrics like precision and F1-score help confirm alignment with business goals \cite{zhang2022conceptualizing} \cite{testi2022mlops}. \textit{Model Training} involves feeding features and algorithms into the model to learn patterns and relationships within the training data, adjusting parameters to minimize prediction errors \cite{sculley2015hidden}. Cross-validation ensures generalization and prevents overfitting \cite{zhang2022conceptualizing} \cite{Lima2022}. Complex models often require GPUs or TPUs, with early stopping and learning rate adjustments to optimize the process \cite{makinen2021needs}. After training, the team will \textit{Evaluate Model \& Responsible AI} utilizes metrics such as ROC-AUC and mean squared error to assess performance on test data \cite{kreuzberger2023machine}. Responsible AI practices ensure fairness by mitigating biases through fairness metrics and continuous monitoring, ensuring ethical and robust model deployment \cite{sculley2015hidden} \cite{zhang2022conceptualizing}. While model development is a core element, it must be supported by an efficient administrative framework and infrastructure setup to ensure smooth operations.

        \begin{table*} [!ht]
            \begin{adjustbox}{max width=\textwidth}
                \begin{tabular}{|>{\centering\arraybackslash}p{2.7cm}|>{\centering\arraybackslash}p{6.7cm}|>{\centering\arraybackslash}p{7.5cm}|}
\hline
\rowcolor{LightCyan}  \textbf{Aspect} & \textbf{MLOps} & \textbf{LLMOps} \\ 
\hline
    Computational Demands
    & Uses standard CPUs or limited GPU resources.
    & Relies heavily on high-performance GPUs or TPUs for both training and inference.
\\ 
\hline
    Cost Structure
    & Driven largely by experimentation, data collection, and initial preprocessing.
    & Dominated by GPU-based inference, high computational costs, and licensing fees for proprietary models.
\\ 
\hline
    Data Management 
    & Requires extensive preprocessing of large-scale, labeled datasets.
    & Involves carefully curated datasets, enabling fine-tuning and effective prompt engineering even with smaller volumes of data.
\\ 
\hline
    Experimentation \& Hyperparameter Tuning
    & Focuses extensively on experimentation to optimize accuracy, performance, and hyperparameter adjustments.
    & Primarily emphasizes fine-tuning foundation models, refining prompts, and balancing performance, computational efficiency, and costs.
\\ 
\hline
    Transfer Learning 
    & Minimal transfer learning; often involves training models from scratch.
    & Heavily depends on transfer learning through fine-tuning of pre-trained foundation models.
\\ 
\hline
    Model Evaluation \& Human Feedback
    & Relies on standard metrics (accuracy, precision, F1-score), with minimal human involvement.
    & Utilizes specialized metrics (e.g., BLEU, ROUGE, DeepEval, FATE), frequently incorporating extensive human feedback, particularly Reinforcement Learning with Human Feedback (RLHF).
\\ 
\hline
    Prompt Engineering 
    & Typically not applicable due to structured data and defined tasks.
    & Crucial for optimizing performance, reducing hallucination, and ensuring reliable, high-quality responses.
\\ 
\hline
    Deployment Complexity \& Latency
    & Simple deployment on standard infrastructure, moderate latency concerns.
    & Requires dedicated GPU infrastructure, complex deployment pipelines (e.g., LangChain, LlamaIndex), and significant latency challenges impacting real-time responsiveness.
\\ 
\hline
    Governance \& Monitoring
    & Governance and monitoring practices are comparatively straightforward, involving standard oversight and regulatory measures.
    & Heightened risks of hallucination, data leakage, and misuse necessitate rigorous, advanced governance frameworks and continuous model monitoring.
\\ 
\hline
\end{tabular}
            \end{adjustbox}
        \caption{Comparison of MLOps and LLMOps derived from Section \ref{sec:mlops_lifecycle}.}
        \label{tab:mlops_llmops_comparison}
        \end{table*}
             
        \textit{Version Control (VC)} is crucial in machine learning for ensuring consistent performance and reproducibility by tracking parameters like data, features, hyperparameters, and source code. Versioning enables reliable model reproduction and maintains experiment integrity and auditability through tools like DVC \cite{dvc} and Git \cite{git_scm} \cite{Lima2022} \cite{kreuzberger2023machine}. Model versions, metadata, training configurations, and performance metrics are also tracked, enabling teams to revert to earlier versions if performance declines, thus preserving system reliability. With version control established, continuous integration (CI) systems help automate testing, ensuring new versions of the model are validated efficiently.
        
        \textit{Continuous Integration (CI)} automates testing and validation of new code and model versions, ensuring system stability. Automated tests, triggered when a new model version is added, assess performance, code quality, and integration issues. CI integrates with version control systems to automatically update code and data, ensuring that all changes are thoroughly tested before merging \cite{9723793} \cite{google-mlops}.

        \begin{table*} [ht]
        \begin{adjustbox}{max width=\textwidth}
        \centering
        \renewcommand{\arraystretch}{1.3}
        \begin{tabular}{|c|c|c|c|c|} \hline 
             \rowcolor{LightCyan} \textbf{MLOps Roles} & \textbf{MLOps Abbreviated Phase} & \textbf{DOL Code} & \textbf{Department Of Labor (DOL) Title} & \textbf{Monthly Wage Estimate DOL 2023} \\ \hline 
             Business Analyst & BN, M & 13-1199 & Business Operations Specialists & \$7,427\cite{bls_oes}\\ \hline 
             System Administrator & AS, VC, D, M & 15-1244 & Network and Computer Systems Administrators &  \$8,381\cite{bls_oes}\\ \hline 
             Network Administrator & AS, VC, D, M & 15-1244 & Network and Computer Systems Administrators &  \$8,381\cite{bls_oes}\\ \hline 
             Data Architect & BN, MD, M & 15-1243 & Database Architects & \$11,419\cite{bls_oes}\\ \hline 
             Data Engineer & BN, MD, M & 15-2051 & Data Scientists & \$9,920\cite{bls_oes}\\ \hline 
             Data Scientist & BN, MD, M & 15-2051 & Data Scientists & \$9,920\cite{bls_oes}\\ \hline 
             ML Engineer & MD, D, M & 15-1299 & Computer Occupations, All Other & \$9,369\cite{bls_oes}\\ \hline 
             Prompt Engineer & MDL, D, M & 15-1299 & Computer Occupations, All Other & \$9,369\cite{bls_oes}\\ \hline
             Software Engineer & AS, MD, VC, D, M & 15-1252 & Software Developers &  \$11,509\cite{bls_oes}\\ \hline 
             QA Engineer & DT, VC & 15-1253 & Software Quality Assurance Analysts & \$9,038\cite{bls_oes}\\ \hline 
             MLOps Engineer & AS, MD, VC, D, M & 15-1252 & Software Developers &  \$11,509\cite{bls_oes}\\ \hline 
             DevOps Engineer & AS, MD, VC, D, M & 15-1252 & Software Developers &  \$11,509\cite{bls_oes}\\ \hline 
             Security and Compliance Officer & BN, AS, MD, VC, D, M & 15-1212 & Information Security Analysts &  \$10,395\cite{bls_oes}\\ \hline 
        \end{tabular}
    \end{adjustbox}
    \caption{MLOps roles mapped to corresponding department of labor codes. The MLOps phase abbreviations can be found in Figure \ref{fig:MLOPS}. Some roles were not an exact match for an existing department of labor occupation so the next closest alternative was selected.}
    \label{tab:Roles}
    \end{table*}

    \begin{table*} [!ht]
    \begin{adjustbox}{max width=\textwidth}
        \renewcommand{\arraystretch}{1.5}
        \centering
\begin{tabular}{|c|c|c|c|c|c|c|c|c|c|c|} \hline
     \rowcolor{LightCyan} \textbf{Tool}                                           & \textbf{Cost/Month} & \textbf{Open Source} & \textbf{BN} & \textbf{AS} & \textbf{MDT} & \textbf{MDL} & \textbf{VC} & \textbf{CI} & \textbf{D} & \textbf{M} \\ \hline
     Delta Lake\cite{delta_lake}, Wekan\cite{wekan}                                       & \$0                 & \checkmark           & \checkmark  &             &              &              &             &             &            &            \\ \hline
     Taiga\cite{taiga}                                                   & \$70                & \checkmark           & \checkmark  &             &              &              &             &             &            &            \\ \hline
     Databricks\cite{databricks}                                              & CB                  & \checkmark           & \checkmark  &             &              &              &             &             &            &            \\ \hline
     Azure ML Studio\cite{azure_ml}                                         & \$0                 &                      & \checkmark  &             & \checkmark   & \checkmark   & \checkmark  &             & \checkmark &            \\ \hline
     AWS Lake Formation\cite{aws_lake_formation}                                      & CB                  &                      & \checkmark  & \checkmark  & \checkmark   & \checkmark   &             &             &            &            \\ \hline
     Asana\cite{asana}                                                   & \$110               &                      & \checkmark  &             &              &              &             &             &            &            \\ \hline
     Clickup\cite{clickup}, Trello\cite{trello}                                         & \$0                 &                      & \checkmark  &             &              &              &             &             &            &            \\ \hline
     Ansible\cite{ansible}, Terraform\cite{terraform}, ARM\cite{azure_rm}, Bicep\cite{azure_bicep}                          & \$0                 & \checkmark           &             & \checkmark  &              &              &             &             &            &            \\ \hline
     Chef\cite{chef} and Puppet\cite{puppet}                                         & CB                  & \checkmark           &             & \checkmark  &              &              &             & \checkmark  &            &            \\ \hline
     PyTorch\cite{pytorch}, Tensorflow\cite{tensorflow}, Pandas\cite{pandas}                             & \$0                 & \checkmark           &             &             & \checkmark   &              &             &             &            &            \\ \hline
     JUnit\cite{junit5}, PyTest\cite{pytest}                                           & \$0                 & \checkmark           &             &             & \checkmark   & \checkmark   &             &             &            &            \\ \hline
     Kubeflow\cite{kubeflow}                                                & \$0                 & \checkmark           &             &             & \checkmark   & \checkmark   &             &             & \checkmark & \checkmark \\ \hline
     Composer by MosaicML\cite{mosaicml2022composer}, H2O.ai\cite{h2oai}                            & CB                  & \checkmark           &             &             & \checkmark   &              &             &             &            &            \\ \hline
     AI Fairness 360\cite{aif360}                                         & \$0                 & \checkmark           &             &             & \checkmark   & \checkmark   &             &             &            & \checkmark \\ \hline
     AWS Sagemaker\cite{aws_sagemaker}                                           & CB                  &                      &             &             & \checkmark   & \checkmark   &             &             & \checkmark & \checkmark \\ \hline
     AWS Kinesis\cite{aws_kinesis}                                             & CB                  &                      &             &             & \checkmark   &              &             &             &            &            \\ \hline
     Neptune.ai\cite{neptuneai}, Weights \& Biases\cite{wandb}                           & \$500               & \checkmark           &             &             & \checkmark   & \checkmark   &             &             &            &            \\ \hline
     LangChain\cite{langchain}, LlamaIndex\cite{llamaindex}, DeepAI\cite{deepai}                           & \$0                 & \checkmark           &             &             &              & \checkmark   &             &             &            &            \\ \hline
     LMStudio\cite{lmstudio}                                                & By Request          & \checkmark           &             &             &              & \checkmark   &             &             &            &            \\ \hline
     Git\cite{git_scm}, DVC\cite{dvc}, FEAST\cite{feast}, Container Registry\cite{docker_hub}, NVIDIA NGC Catalog\cite{nvidia_ngc} & \$0                 & \checkmark           &             &             &              &              & \checkmark  &             &            &            \\ \hline
     Bitbucket\cite{bitbucket}                                               & \$33                &                      &             &             &              &              & \checkmark  &             &            &            \\ \hline
     Azure Container Registry\cite{azure_container_registry}                                & \$5                 &                      &             &             &              &              & \checkmark  &             &            &            \\ \hline
     MLFlow\cite{mlflow}                                                  & \$0                 & \checkmark           &             &             & \checkmark   & \checkmark   & \checkmark  & \checkmark  &            &            \\ \hline
     Hugging Face\cite{huggingface}                                            & \$0                 & \checkmark           &             &             & \checkmark   & \checkmark   & \checkmark  &             & \checkmark &            \\ \hline
     Azure DevOps\cite{azure_devops}                                            & \$30                &                      & \checkmark  &             &              &              & \checkmark  & \checkmark  & \checkmark &            \\ \hline
     GitLab Enterprise\cite{gitlab_enterprise}                                       & \$290               & \checkmark           & \checkmark  &             &              &              & \checkmark  & \checkmark  & \checkmark &            \\ \hline
     OpenLLM\cite{openllm}, TensorFlow Serving\cite{tensorflow_serving}                             & \$0                 & \checkmark           &             &             &              &              &             &             & \checkmark &            \\ \hline
     Jenkins\cite{jenkins}                                                 & \$0                 & \checkmark           &             &             &              &              &             & \checkmark  &            &            \\ \hline
     Seldon\cite{seldon}                                                  & \$0                 & \checkmark           &             &             & \checkmark   &              & \checkmark  &             & \checkmark & \checkmark \\ \hline
     ELK Stack\cite{elastic_stack}, Prometheus\cite{prometheus}, Grafana\cite{grafana}, OpenLLMetry\cite{traceloop_openllmetry}             & \$0                 & \checkmark           &             &             &              &              &             &             &            & \checkmark \\ \hline
     Grafana Cloud\cite{grafana_cloud}                                           & \$19                & \checkmark           &             &             &              &              &             &             &            & \checkmark \\ \hline
     Whylabs\cite{whylabs}                                                 & \$250               & \checkmark           &             &             &              &              &             &             &            & \checkmark \\ \hline
     Datadog\cite{datadog}                                                 & CB                  & \checkmark           &             &             &              &              &             &             &            & \checkmark \\ \hline
     Langsmith\cite{langsmith}                                               & \$390               & \checkmark           &             &             &              & \checkmark   & \checkmark  &             &            & \checkmark \\ \hline
\end{tabular}
    \end{adjustbox}
    \caption{Tools that are used in each MLOps lifecycle phase along with their monthly cost. The MLOps phase abbreviations can be found in Figure \ref{fig:MLOPS}. Consumption based pricing (CB) is for services priced by consumption instead of subscription. The cost per months is based on 10 users.}
    \captionsetup{format=plain}
    \label{tab:tools-to-phases}
    \end{table*}
        
        \textit{Model Deployment (D)} involves comprehensive testing, validation, and monitoring before and after production deployment. In the \textit{Test} envinronment, the model is tested in production-like conditions to validate its performance. Quality assurance tests, including batch and near real-time inference, assess large data processing and time-sensitive accuracy. Responsible AI checks prevent bias, while integration tests detect critical issues early, ensuring the model’s reliability before deployment \cite{9355312}. \textit{Gated Approval} in MLOps ensures approved model deployment by requiring authorization before models progress to staging or production. Requiring gated approval reduces the risk that an individual pushes software and resources into production before they are fully vetted and approved by the appropriate personnel. In \textit{Staging/Production}, once the model passes quality assurance tests, the entire application is deployed for end-user access, supporting both batch and near real-time inferencing. Performance benchmarks, such as the MLPerf Inference Benchmark presented in Reddi's work (2020), can be used to evaluate both inference types, confirming production readiness, ensuring smooth transitions, and preserving accurate predictions. \cite{9138989}.
             
        Continuous \textit{Monitoring (M)} aggregates insights from test, staging, and production environments to track performance metrics. This system enables adaptive responses to operational issues, especially in multi-tenant environments. Any deviations trigger adjustments to ensure that the model continues to meet business requirements \cite{6740239}.

    In summary, MLOps lifecycle framework establishes a comprehensive, scalable system for managing machine learning lifecycles, facilitating smooth transitions between development, deployment, and monitoring. The integration of tools such as version control, CI/CD pipelines, and responsible AI ensures operational efficiency and ethical model performance.

\section{MLOps Roles and Tools}
The previous section describes a breakdown of each phase and the associated tasks in the MLOps lifecycle framework. We now turn our focus to the roles and tools that are involved. The roles, emphasized in italllics, can be found in Table \ref{tab:Roles}. The comprehensive list of tools are presented in Table \ref{tab:tools-to-phases}.

The first priority is to align business needs and Key Performance Indicators (KPIs) with the effort to build and deploy a model. It's the job of the \textit{Business Analysts} \cite{bls_oes} to focus on the strategic side of MLOps. They align the MLOps process with business objectives by translating business needs into epics and milestones using tools like Azure DevOps, Wekan \cite{wekan} or Taiga \cite{taiga}. The rest of the team will break down these large long term goals into smaller short term technical goals. Given the established business strategy, \textit{System Administrators} and \textit{Network Administrators} \cite{bls_oes} will design, create and maintain the cloud and on prem infrastructure using tools like Ansible \cite{ansible}, Terraform \cite{terraform} and Bicep \cite{azure_bicep}. Once the infrastructure is created and proper access has been verified by the team, \textit{Data Architects} and \textit{Data Engineers} \cite{bls_oes} can get to work on ingesting, transforming, storing and versioning the data. Databricks \cite{databricks} and Azure ML Studio \cite{azure_ml} are examples of tools that both Data Engineers and \textit{Data Scientists} use but Data Scientists will also analyze, visualize and build statistical models with Python and R. \textit{ML Engineers} can then take the prepared data to create and/or fine tune models with PyTorch \cite{pytorch}, Tensorflow \cite{tensorflow}, Azure ML Studio and MLFlow \cite{mlflow}. Rather than building a model from scratch, a \textit{Prompt Engineer} could design a LLM prompt to achieve the desired output. Training and tuning iterations are stored in MLFlow and models are stored in repositories like Azure ML Studio and Huggingface. \textit{Software Engineers} then develop applications that interact with the models to deliver value to customers using their favorite Integrated Development Environment(IDE) and versioning their source code with Git or Bitbucket. A \textit{QA Engineer} \cite{bls_oes} will use testing frameworks like JUnit  \cite{junit5} or PyTest \cite{pytest} to ensure the software performs as defined in the feature request. At this point everything is tied together by \textit{DevOps Engineers} and \textit{MLOps Engineers} \cite{bls_oes} to ensure the software and models are built, tested and deployed onto the prepared infrastructure in a reliable and reproducible manner. The appropriate model version is paired with a compatible source code version and built into a deployable image that can be stored in a Container Registry and deployed into the cloud or into a Kubeflow cluster. The deployment pipelins should also ensure automated configuration of logging, traces and metrics for monitoring. Finally, \textit{Security and Compliance Officers} play a key role in ensuring that the entire process adheres to legal standards and remains protected from unauthorized access.

The creation of an MLOps pipeline involves a variety of tools across different phases of the machine learning lifecycle framework, ranging from data collection to model development to deployment and monitoring. Each tool plays a critical role in ensuring that the entire process runs efficiently, and the flexibility in choosing tools is key to customizing the pipeline based on team skills, infrastructure, and specific project needs.

\section{Conclusion}
This paper covers several existing MLOps maturity levels and proposes a consolidated lifecycle framework that includes LLMOps along with the roles and tools involved. A consolidated framework creates a shared understanding and common terminology, building on lessons learned from similar challenges. Mapping roles and tools to MLOps phases aids in cost estimation and resource allocation. As teams adopt MLOps practices, the deployments of machine learning applications become more reliable, reproducible, scalable, and observable. The lifecycle framework also serves as a foundation for iterative improvement as MLOps principles evolve. However, there are several areas that could benefit from additional research such as
\begin{enumerate}
    \item A case study to apply the framework to a new AI-enabled application project, tracking implementation time, costs, and tools involved.
    \item Review and highlight improvements to our framework, particularly with the integration of LLMOps.
    \item A comprehensive survey of the threat landscape that identifies potential attacks and challenges in securing MLOps, while also recommending strategies for mitigating these threats.
    \item Determine safeguards to protect AI-integrated edge devices, such as security cameras, autonomous vehicles, and healthcare devices, from data poisoning and cyber threats?
\end{enumerate}
Although this paper addresses several improvements, investigating security implications while implementing the MLOps lifecycle framework is an area that would benefit from future research. As organizations strive to achieve more reliable and consistent model deployments, this work builds on the foundation of other MLOps research and paves the way for future advancements.

\section*{Acknowledgment}
This work was supported by the Predictive Analytics and Technology Integration (PATENT) Laboratory at the Department of Computer Science and Engineering, Mississippi State University.

\bibliographystyle{ieeetr} 
\bibliography{sources.bib,sources.web}

\end{document}